\documentstyle[twocolumn,floats,prl,aps,psfig]{revtex}
\tighten

\begin{document}

\title{Effective Restoration of the $U_A(1)$ symmetry in the SU(3) linear
  $\sigma$ model}

\author{J\"urgen Schaffner-Bielich}

\address{RIKEN BNL Research Center, Brookhaven National Laboratory,
Upton, New York 11973-5000, USA}

\date{\today}
\maketitle

\begin{abstract}
The effective restoration of the chiral $U_A(1)$ symmetry in
strong interactions is studied using the linear chiral SU(3)$\times$SU(3)
model at finite temperatures. We find that the disappearance of the chiral anomaly
causes a considerable change in the meson mass spectrum.
We propose several signals for detecting this chiral phase
in ultrarelativistic heavy-ion collisions:
The $\eta/\pi^0$ ratio is enhanced by an order of magnitude,
the $a_0$ is suppressed in the $K\bar K$ mass spectrum, and 
the scalar $\kappa$ meson appears as a peak
just below the $K^*(892)$ in the invariant $\pi K$ mass spectrum.
\end{abstract}

\draft


\vspace*{-.5cm} 
Probing the phase transition of strongly interacting matter at finite
temperature is one of the primary goals of ultrarelativistic heavy-ion
collisions \cite{QM99}. At BNL's Relativistic Heavy-Ion Collider (RHIC) 
the central collision zone of the two gold ions may create temporarily hot
matter, presumably similar to what is studied numerically in lattice gauge
calculations. Present simulations 
indicate that the phase transition seen in hot quantum chromodynamics (QCD)
coincides with the chiral phase 
transition \cite{Laer98}. But the fundamental question of which chiral symmetry is restored
in hot QCD is not settled yet. 
There are two scenarios according to Shuryak \cite{Shuryak94}: 
only flavor SU(3) chiral symmetry is restored or both SU(3) and $U_A(1)$ symmetries
are restored.
The latter scenario is particularly interesting, as it might lead to a first order
phase transition for two-flavor QCD and drastic effects 
for the spectrum of pseudoscalar mesons \cite{Pisa84}.

The $U_A(1)$ symmetry is broken by nonperturbative
effects \cite{Hooft76} which gives the $\eta'$ a finite mass in the chiral limit.
The topological susceptibility in pure color SU(3) gauge theory can be
linked with the finite $\eta'$ mass via the Veneziano-Witten formula
\cite{Witten79,Vene79}. 
The restoration of the $U_A(1)$ symmetry 
should then be visible in the behavior of the topological susceptibility around
the critical temperature. 
There exists some theoretical estimates in the
literature how the chiral
anomaly behaves at finite temperatures.
Changes in the topological susceptibility 
should be small for low temperatures \cite{Shuryak94b}.
In the plasma phase above $T_c$ it should drop exponentially due to
instanton effects \cite{Pisa80}. More dramatically, it was shown in
\cite{Kharzeev98} that it even vanishes at $T_c$ in the large $N$ limit.

The topological susceptibility has been studied on the lattice, both in pure
color SU(3) gauge theory \cite{Chu95,Alles97} as well as for the unquenched case
\cite{Chu97,Alles99}. In both cases, a sharp drop by an order of magnitude
is observed at the deconfining temperature
showing the apparent restoration of the $U_A(1)$ symmetry. 
Other approaches utilize chiral susceptibilities,
where simulations indicate that the $U_A(1)$ symmetry
is not completely restored \cite{Bernard97} but its effects are at or below
the 15\% level \cite{Chandra99}. 
The screening mass difference of the pion and the $a_0(\delta)$ have been also 
studied as a measure of $U_A(1)$ breaking effects. 
Using staggered fermions it has been found that 
the anomalous chiral symmetry is probably not
fully restored at $T_c$  but that
the $a_0-\pi$ mass splitting drops drastically \cite{Laer96,Gottlieb97,Kogut98}.
Simulations with domain wall fermions   
demonstrate that anomalous chiral symmetry breaking effects 
are at or below the 5\% level above but close to $T_c$ \cite{Chen98}. 
Note that all the above results consistently indicate that
effects from the $U_A(1)$ breaking are strongly suppressed, i.e.\  
an {\em effective} restoration of the $U_A(1)$ symmetry close to $T_c$.  

Recently, the issue of finding signals for the restoration of chiral symmetry
in ultrarelativistic heavy-ion collisions has received considerable
attention. E.g.\ signals for the restoration of the SU(2) chiral symmetry associated 
with the $\sigma$ meson have been
proposed in \cite{Song97,Chiku98}. 
In particular, signals for the partial
restoration of the $U_A(1)$ symmetry in connection with the $\eta'$ meson have
been invoked in \cite{Kharzeev98,Kapusta96,Huang96,Vance98}. 
Effects of the $U_A(1)$ anomaly on meson masses 
have been also studied within the SU(3) Nambu--Jona-Lasinio model by Kunihiro
\cite{Kuni89,Kuni91}.

Here, we are going to study the full SU(3) linear $\sigma$ model at finite
temperature including effects from the effective restoration of the $U_A(1)$
symmetry. Our aim is to extract signals from the strong suppression of
$U_A(1)$ breaking effects in a chirally restored phase.
The additional scalar mesons in the SU(3) $\sigma$ model compared to the
standard SU(2) $\sigma$ model 
provides novel signals for the effective restoration of the $U_A(1)$ symmetry: 
there will be an enhancement of the number of $\eta$ mesons due to a feedback
from the decay of $a_0$ mesons, its chiral partner, which will be suppressed
then in the $K\bar K$ mass spectrum, and there appears a new
scalar resonance in the invariant $\pi K$ mass spectrum, stemming from the chiral partner of
the kaon, the $\kappa$ meson.

The inclusion of the strangeness degree of freedom, i.e.\ going from
SU(2) to SU(3), finds its justification
in the recent findings that the strange quark mass is about $m_s=100-150$  
MeV \cite{Blum99}, close to half the expected temperatures at RHIC of
$T\approx 200-300$ MeV in the initial stage of the collision. Therefore, strangeness
has to be included in the linear $\sigma$ model for studying hot matter
relevant to physics at RHIC.

The SU(3) linear $\sigma$ model is known for a long time \cite{Levy67}. Only
recently, more than thirty years later, there has been a renaissance 
of this chiral Lagrangian. The resurrection of the $\sigma$
meson \cite{Torn96} and the finding of the $\kappa(900)$ resonance in the $\pi K$
scattering data \cite{Ishida97,Black98,Oller99} leads to the conclusion that
there is a low-mass scalar nonet \cite{Black98b}.
Jaffe predicted this scalar nonet long ago \cite{Jaffe77a} 
as being built out of $(q\bar q q\bar q)$ states with an inverted mass spectrum
compared to the pseudoscalar meson spectrum.
Recent lattice simulations are supporting this picture \cite{Alford00}. 
It is interesting to note that the inversion of the mass spectrum is implemented 
in the SU(3) linear $\sigma$ model by the anomaly term which has the opposite
sign for the pseudoscalar and scalar masses.  
The model Lagrangian gives a surprisingly good agreement with the
data already at tree-level \cite{Ishida98,Torn99}.
The model was also extended to
finite temperatures \cite{Meyer96,SR99} without implementing 
effects from the effective $U_A(1)$ symmetry restoration. 

Let us now write down the SU(3)$\times$SU(3) chiral Lagrangian:
\begin{eqnarray}
\lefteqn{
  {\cal L} = \frac{1}{2} {\rm Tr} \left(\partial_\mu \Phi^\dagger \partial^\mu \Phi
  + \mu^2 \Phi^\dagger \Phi \right)
  - \lambda {\rm Tr} \left( \Phi^\dagger \Phi\right)^2} \nonumber \\
&&{}  - \lambda' \left( {\rm Tr} \Phi^\dagger \Phi\right)^2 
+ c \cdot \left( \det \Phi + \det \Phi^\dagger \right) + \epsilon \cdot
\sigma + \epsilon' \cdot \zeta \;.
\end{eqnarray}
where $\Phi$ is a 3x3 complex matrix describing the pseudoscalar and scalar nonet.
The term proportional to the determinant  breaks the $U_A(1)$ symmetry. The last two
terms break chiral symmetry explicitly, which are simulating effects from a finite light quark
and a strange quark mass, respectively. 
There are two order parameters corresponding to the light quark condensate
$\sigma$ and the strange quark condensate $\zeta$.
Chiral symmetry is restored at a certain temperature as the $\sigma$ order parameter
drops towards zero (and effects from the explicit symmetry breaking term
$\epsilon\cdot\sigma$ can be ignored). Then the masses of the
pion and the sigma mesons and the masses of the $\eta$ and the $a_0$ meson are 
the same separately.
Their mass gap is proportional to the coefficient $c$ of the anomaly term times
the strange order parameter $\zeta$:
\begin{equation}
  m_\pi = m_\sigma <  m_{a_0} = m_\eta \quad , \qquad \Delta m = 4 c\cdot \zeta
\quad .
\end{equation}
If the chiral $U_A(1)$ symmetry is effectively restored, then this mass gap
vanishes and all four meson masses are the same irrespective of the value of $\zeta$:
\begin{equation}
 m_\pi = m_\sigma \approx m_{a_0} = m_\eta \quad \mbox{ for } c\approx 0  
\quad .
\end{equation}
Note that this is only the case for a (effective) restoration of the $U_A(1)$
symmetry ($c\approx 0$) as the strange order parameter $\zeta$ will not decrease
strongly at $T_c$ due to the finite strange quark mass \cite{Kogut91}.

We assume now that the coefficient of the anomaly term $c$ drops
as a function of temperature. As a guideline we take the temperature effects
to be proportional to the topological susceptibility in pure glue theory as suggested by the
Witten-Veneziano formula \cite{Witten79,Vene79}. We use the lattice data as published
in \cite{Alles97} so that the coefficient is nearly constant until $T_c$ and
drops then by an order of magnitude (but is not vanishing). We take the
critical temperature to be $T_c=150$ MeV as deduced from recent investigations 
on the lattice (see \cite{Laer98} for a summary). 
Thermal excitations for all pseudoscalar and
scalar fields are taken into account in a
Hartree scheme in a selfconsistent way. The gap
equations for the two order parameters are solved together with the
expressions for the meson masses and the thermal excitations iteratively until 
convergence is achieved. 

\begin{figure}[t]
\psfig{file=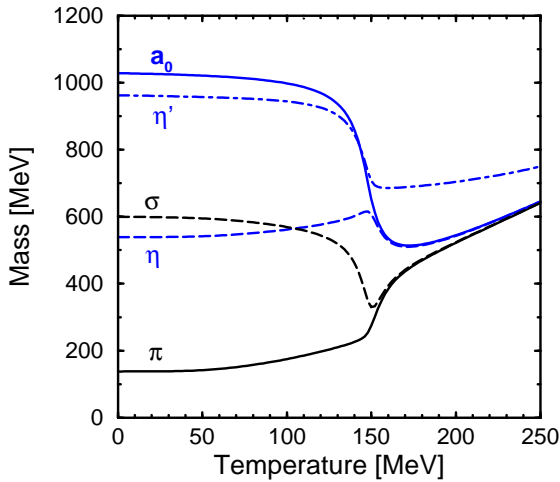,width=0.41\textwidth}
\caption{Meson masses versus temperature for the $\pi$, $\sigma$, $\eta$,
  $\eta'$ and $a_0$. There is a level crossing of the $\eta$ and $\eta'$
  states. The decay $a_0\to\pi+\eta$ is blocked just below the
  chiral phase transition.} 
\label{fig:masses1}
\end{figure}

\begin{figure}[t]
\psfig{file=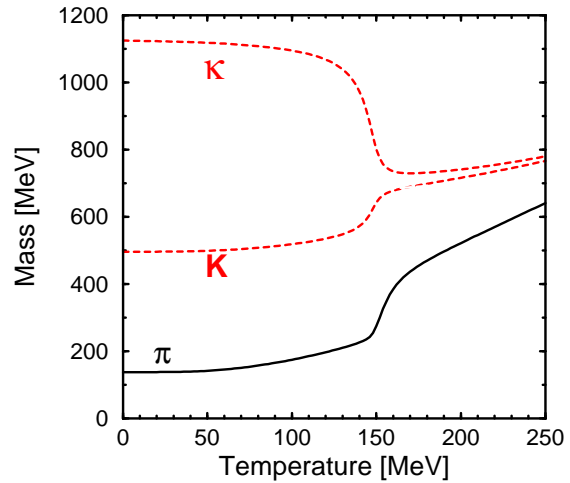,width=0.41\textwidth}
\caption{Meson masses versus temperature for the pion, kaon,
  and $\kappa$. The decay $\kappa\to\pi + K$ is blocked just
  below~$T_c$.}
\label{fig:masses2}
\end{figure}

The meson masses as a function of temperature are shown in Figs.\
\ref{fig:masses1} and \ref{fig:masses2}. The most striking feature compared to 
a calculation with a constant anomaly coefficient is that the phase transition 
is shifted to lower temperatures. 
For a constant coefficient, the phase transition is much less
pronounced and happens at $T_c\approx 210$ MeV \cite{SR99}. Here,
the chiral phase transition happens precisely at the same temperature at which 
the anomaly coefficient is put to drop down, i.e.\ at $T_c=150$ MeV. 
Remarkably, the restoration of $U_A(1)$ symmetry shifts the value of the critical
temperature and strengthens the chiral phase transition which was also seen
within the Nambu--Jona-Lasinio model \cite{Kuni89}.
The $U_A(1)$ symmetry is restored at $T\approx 250$ MeV
which can be read off of Fig.\ \ref{fig:masses1} by the degeneracy 
of the chiral partners $\pi-\sigma$ and $\eta-a_0$. 
The overall meson mass spectrum changes significantly across the phase
transition. The mass of the $\eta'$ drops down considerably to
$m_{\eta'}\approx 650$ MeV 
and is then approximately degenerate with the kaon mass above $T_c$. 
The pseudoscalar mixing angle is ideal above $T_c$ due to the smallness of the
anomaly term. This means in principle that the $\eta'$, as the chiral partner of the $a_0$,
is a purely light quark system while the $\eta$ becomes purely strange.
Nevertheless, it is apparent from Fig.\ \ref{fig:masses1} that 
a level crossing of the $\eta$ and the $\eta'$ masses occurs around $T_c$
(see also \cite{Pisa84,Kuni89}).
Hence, the $\eta$ and $\eta'$ are
switching their identity at $T_c$. The $\eta$ is now the chiral partner of the 
$a_0$ and is the nonstrange state while the $\eta'$ is the pure strange quark
state. 
Surprisingly, the $\eta'$ mass is even slightly smaller than the single strange quark state, 
the kaon, which is an interesting nontrivial effect originating from the
different thermal contributions for the kaon and the $\eta_s$ masses.
The number of $\eta'$ mesons will then be enhanced in 
the hot medium. In return, the number of $\eta$ mesons
will increase then by the decay $\eta'\to\eta\pi\pi$ at freeze-out 
which will be seen by a modified slope parameter in the low transverse
momentum spectra.  

A more dramatic effect i s associated with the scalar isovector meson $a_0$.
The free $a_0$ decays mainly to $\eta+\pi$ and to two kaons. The decay to two
pions is forbidden as it violates isospin.
As evident from Fig.\ \ref{fig:masses1},  the $a_0$ mass decreases strongly
with temperature, as is it also seen on the lattice
\cite{Laer96,Gottlieb97,Kogut98,Chen98}. 
As chiral SU(3) symmetry is 
restored at $T_c$ , its mass gets degenerate with the $\eta$ mass, its chiral partner.
In addition, as $U_A(1)$ is effectively restored, the $a_0$ and $\eta$ masses
will be close to the pion mass just above $T_c$ (as seen in Fig.\
\ref{fig:masses1}). 
Hence, already {\em below} $T_c$ the decay
$a_0\to\eta + \pi$ must be blocked by phase space just when the mass
difference of the $a_0$ and the $\eta$ equals the thermal pion mass.
Also, the decay matrix element of $a_0\to\eta +\pi$
is considerably reduced above $T_c$ as it is proportional to the $\sigma$
order parameter and the anomaly term.
The decay to two
kaons is heavily suppressed as the $a_0$ is actually lighter than one kaon
alone if $U_A(1)$ is effectively restored (see Fig.\ \ref{fig:masses1} compared 
with Fig.\ \ref{fig:masses2}). 
Hence, the inelastic channels for the $a_0$ are closed above $T_c$. 
The elastic channels are still large as
they are proportional to the coupling constant $\lambda$ which is of the order 
of 10. We conclude that there is
chemical equilibrium between the $a_0$, $\eta$, $\sigma$ and pion in the
chiral phase as it is U(2)$\times$U(2) symmetric.
This argument is supported by the work of Song and Koch \cite{Song97} who find 
that the $\sigma$ and pion mesons are in chemical equilibrium
in the SU(2) linear $\sigma$ model.

For detecting this chiral symmetric phase in ultrarelativistic heavy--ion collisions,
the expansion stage of the hot and chemically equilibrated matter must be
short and/or out of equilibrium.
If the expansion is adiabatically,
the system adjusts itself, freezes out like a free gas and no effect will be visible.
A signal will show up, if either the system expands 
from the chiral phase above $T_c$ until the freeze-out
temperature $T_f$ faster than the lifetime of the $a_0$ of about $2-4$ fm or
if the chemical freeze-out happens before the thermal freeze-out \cite{Huang96}.
Then the number of $a_0$'s is approximately conserved during the
expansion and the numbers of $a_0$'s at freeze-out will be approximately equal
to the numbers of pions and three times the number of $\eta$ mesons (due
to isospin counting). The $a_0$ will then mainly decay to $\eta+\pi$
increasing the numbers of the $\eta$ mesons drastically. 
Taking into account the decays $a_0\to\eta +\pi$ and $\sigma\to2\pi$, we get the ratio
$\eta/\pi^0=(3n_{a_0}+n_\eta)/(n_{a_0}+2/3n_\sigma+n_{\pi^0})=3/2$ as a signal 
of the formation of an U(2)$\times$U(2) symmetric phase. 
Hence, it is even possible to produce more $\eta$
mesons than $\pi^0$'s. 
The two-kaon decay channel is suppressed at finite temperature
due to the larger mass of the kaon
and the smaller mass of the $a_0$ at $T_f$ \cite{Horst99}. This decay is a
subthreshold process even at $T=0$ so that a slight change in the masses will 
reduce the branching ratio.

\begin{figure}[t]
\psfig{file=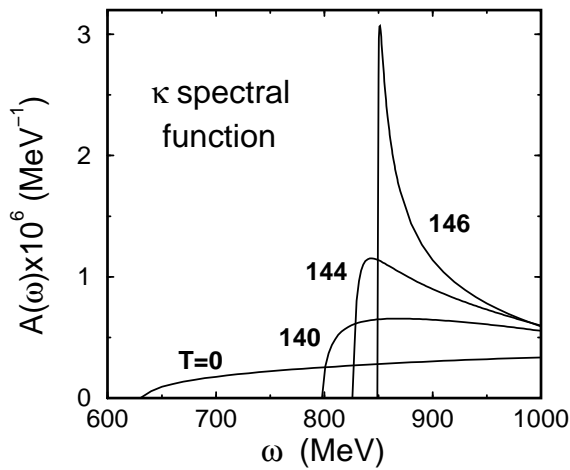,width=0.41\textwidth}
\caption{The spectral function for the $\kappa$ meson for temperatures of
  $T=0,140,144,146$ MeV.  A pronounced threshold enhancement appears for
  temperatures close to $T_c$.} 
\label{fig:spectral}
\end{figure}

We discuss now another observable related to the effective
restoration of the $U_A(1)$ which is associated with the $\kappa(900)$ resonance.
The $\kappa$ meson is very broad similar to the $\sigma$ meson
\cite{Ishida97,Black98,Oller99} and decays to a pion and a kaon.
Its mass depends strongly on the $U_A(1)$ anomaly and is decreasing with
temperature as shown in Fig.\ \ref{fig:masses2}.
The decay width depends on the coefficient of the anomaly term
and decreases therefore in the chiral $U_A(1)$ phase.
We find at tree-level that the width changes from values around $\Gamma\approx .8$
GeV to $\Gamma\approx .2$ GeV when setting the contribution from the anomaly
term to zero. Hence, the barely visible broad resonance gets a 
much smaller width in the chiral $U_A(1)$ phase and can possibly be seen. 
As indicated by Fig.\ \ref{fig:masses2}, the mass of the $\kappa$
approaches that of the kaon towards chiral symmetry restoration so that the
strong decay $\kappa\to K+\pi$ is blocked by 
phase space. This happens already below $T_c$ similar as for the $a_0$.
The $\kappa$ meson can then be visible in the invariant $\pi K$ mass 
spectrum, if the system freezes out dominantly around $T_c$.
Chiku and Hatsuda have demonstrated this effect in connection
with the $\sigma$ meson appearing in the $\pi\pi$ channel \cite{Chiku98}.
The decay channel $\kappa\to \pi+K$ opens just below $T_c$ so that there will be a
pronounced cusp structure in the corresponding spectral function as depicted
in fig.\ \ref{fig:spectral}.
The $\kappa$ resonance will then emerge in the $\pi K$ invariant mass spectrum 
around 850 MeV.
For the $\pi K$ mass spectrum, there is a only one background in that mass
region, which is from the vector kaon, the $K^*(892)$. Vector
meson masses go up with temperature if studied in a SU(2) gauged chiral
Lagrangian \cite{Pisa95} as they have to be degenerate with their
heavier axial vector chiral partners. 
A study in a SU(3) linear $\sigma$ model with vector mesons
shows that the $K^*(892)$ mass
stays approximately constant until $T_c$ and effectively rises then for higher
temperatures \cite{Ziesche99}. Hence, the background from $K^*(892)$
should be comparably low.

The appearance of
the $\kappa$ in the $\pi K$ spectra can be 
detected by two particle correlation at BNL's RHIC. Here, the same
techniques employed for reconstructing the $\rho$ in the $\pi\pi$ mass spectrum
\cite{Jack99} can be utilized. 
The STAR group, as well as BRAHMS and PHOBOS, will reconstruct $\phi$ mesons
in the $K \bar 
K$ spectra \cite{Fuqiang} where the $a_0$ will be seen, too.
The $\eta$ meson will be measured
in the diphoton spectra at the PHENIX detector \cite{Mike} and the enhancement 
proposed here can be checked.

We are indebted to Miklos Gyulassy for initiating this work, Dima Kharzeev for
illuminating discussions about the chiral anomaly, and Larry McLerran for
critical comments. We thank Jonathan Lenaghan, 
Shigemi Ohta, Rob Pisarski, Dirk Rischke, Shoichi Sasaki, Thomas Sch\"afer, Edward
Shuryak, and Horst St\"ocker for helpful conversations.


\end{document}